\documentclass[preprint]{aastex}

\begin{document}

\title{The Tolman Surface Brightness Test for the Reality of the Expansion. 
III. HST Profile and Surface Brightness Data for Early-Type Galaxies in
Three High-Redshift Clusters}

\author{Lori M. Lubin\altaffilmark{1,2}} \affil{Department of Astronomy,
California Institute of Technology,\\ Mailstop 105-24, Pasadena,
California 91125}

\altaffiltext{1}{Hubble Fellow} \altaffiltext{2}{Current address :
Department of Physics and Astronomy, Johns Hopkins University,
Baltimore, MD 21218}

\author{Allan Sandage} 
\affil{Observatories of the Carnegie Institution of Washington,\\ 813
Santa Barbara Street, Pasadena, California 91101}

\begin{abstract}

Photometric data for 34 early-type galaxies in the three high-redshift
clusters Cl 1324+3011 ($z = 0.76$), Cl 1604+4304 ($z = 0.90$), and Cl
1604+4321 ($z = 0.92$), observed with the {\it Hubble Space Telescope}
(HST) and with the Keck 10-meter telescopes by Oke, Postman \& Lubin,
are analyzed to obtain the photometric parameters of mean surface
brightness, magnitudes for the growth curves, and angular radii at
various Petrosian $\eta$ radii. The angular radii at $\eta = 1.3$ mag
for the program galaxies are all larger than 0\farcs{24}. All of the
galaxies are well resolved at this angular size using HST whose
point-spread function is 0\farcs{05}, half width at half maximum. The
data for each of the program galaxies are listed at $\eta$ = 1.0, 1.3,
1.5, 1.7, and 2.0 mag. They are corrected by color equations and $K$
terms for the effects of redshift to the rest-frame Cape/Cousins $I$
for Cl 1324+3011 and Cl 1604+4304 and $R$ for Cl 1604+4321. The $K$
corrections are calculated from synthetic spectral energy
distributions derived from evolving stellar population models of
Bruzual \& Charlot which have been fitted to the observed broad-band
($BVRI$) AB magnitudes of each program galaxy.  The listed photometric
data are independent of all cosmological parameters. They are the
source data for the Tolman surface brightness test made in Paper IV.

\end{abstract}


\section{Introduction}

In the first two papers of this series (Sandage \& Lubin 2001,
hereafter Paper I; Lubin \& Sandage 2001a, hereafter Paper II), we
review the Tolman (1930, 1934) surface brightness test for the reality
of the expansion. In Paper I, we present the calibration of the
diagnostic diagrams at zero redshift of mean surface brightness
$\langle SB \rangle$, absolute magnitude $M$, and linear radius R at
various Petrosian (1976) $\eta$ metric radii. The Petrosian $\eta$
function is defined as the difference in magnitude between the mean
surface brightness averaged over the area interior to a particular
radius and the surface brightness at that radius. The unique
properties of this function in defining a metric size, which is
essential when comparing parameters of galaxies at different
redshifts, are described in detail in \S2 of Paper I.  For the local
calibrations, we have used the photometric data of Postman \& Lauer
(1995) for first ranked cluster galaxies in 118 low redshift Abell
clusters. In Paper II, we describe the effect on the profile data of
finite angular resolution and the differences in the diagnostic
diagrams for early-type galaxies of different flattening ratios.
         
The purpose of the present paper is to measure the three galaxy
parameters of mean surface brightness $\langle SB \rangle$, apparent
magnitude $m$, and angular radius $r$ as a function of the Petrosian
radius $\eta$ for early-type galaxies at high redshift. We have
obtained this galaxy sample from the three high-redshift clusters Cl
1324+3011 ($z = 0.76$), Cl 1604+4304 ($z = 0.90$), and Cl 1604+4321
($z = 0.92$). These clusters were observed with the {\it Hubble Space
Telescope} (HST) by Postman, Lubin, and Oke as discussed by Oke,
Postman \& Lubin (1998, hereafter OPL), Postman, Lubin \& Oke (1998,
2001; hereafter, PLO98 and PLO01), and Lubin et al. (1998, 2001). We
have used their extensive photometric data for the present paper.  The
sample selection and the methods of calculating the $\langle SB (\eta)
\rangle$, $m(\eta)$, and $r(\eta)$ functions for these galaxies are
given in \S 2. The method of calculating the photometric $K$
corrections in the Cape/Cousins $RI$ photometric system is given in \S
3. The surface brightnesses and apparent magnitudes in $I$ for the
clusters Cl 1324+3011 and Cl 1604+4304 have been corrected by the
$K(I)$ correction to the rest-frame $I$ system.  The data in $R$ for
Cl 1604+4321 have been corrected by $K(R)$ to the rest-frame $R$
system. These data listed in Tables 5--7 are free from all
cosmological assumptions as to world model; however, for the analysis
in Lubin \& Sandage (2001b, hereafter Paper IV) the absolute magnitude
($M$) and linear radius (R) values, each which require knowledge of
the deceleration parameter $q_o$, are needed for the comparison of the
high-redshift data with the calibrations of the diagnostic diagrams of
Paper I.  Because the basic observational data in the present paper
are cosmological-model free, whereas the analysis in Paper IV is not,
we have separated the discussions into the two papers (III here, and
IV) to emphasize the difference between the data source presented in
this paper with its analysis in Paper IV.

\section{Measurements of Mean Surface Brightness, Apparent Magnitude,
and Radius for the High-Redshift Early-Type Galaxies}

\subsection{The Color and Magnitude System} 
\subsubsection{The Keck Observations} 

The observations by Oke, Postman \& Lubin (1998) of the three clusters
were made both with the Keck 10-meter telescopes and with HST. The CCD
observations at Keck were made in the four photometric passbands that
are close to, but not identical with, the standard $BV$ system of
Johnson and the $RI$ system of Cape/Cousins (see Figure 1 of OPL).
The data on the natural Keck system were reduced to the standard
$BVRI$ systems by observing selected standard stars measured by
Landolt (1992). These were chosen to be adequately faint in order not
to saturate in the short Keck exposures that were taken to determine
the color equations.

The importance of the Keck $BVRI$ data for these high-redshift
galaxies is that the extended color range permits the choice of an
appropriate synthetic spectral energy distribution (SED) of each
galaxy from the library of evolved SED models of Bruzual \& Charlot
(1993). These evolved rest-frame SEDs have been used to calculate the
$K(R)$ and $K(I)$ corrections in \S 3.

\subsubsection{The HST Observations}

The profile data needed for the calculation of the mean surface
brightness $\langle SB \rangle$, apparent magnitude $m$, and angular
radius $r$ at various $\eta$ values were obtained using HST for the
same galaxies observed at the Keck telescopes. Clusters Cl 1324+3011
($z = 0.76$) and Cl 1604+4304 ($z = 0.90$) were observed using the
Wide Field Planetary Camera 2 (WFPC2) in the $F814W$ bandpass, while
cluster Cl 1604+4321 ($z = 0.92$) was observed in the $F702W$
bandpass. The color and magnitude equations between these HST filter
systems and the standard Cape/Cousins ($RI$) systems are

\begin{equation}
           R = F702W + 0.272 (V-R),
\end{equation}

\noindent and 

\begin{equation}
     I = F814W - 0.05~{\rm for~all}~(V-I) > 0.7.
\end{equation}

\noindent These equations are the linear approximations to the
quadratic equations of Holtzman et al.\ (1995). They are accurate to
better than 0.01 mag.

The sense of equation (1) is that the $R$ magnitudes are fainter than
the $F702W$ magnitudes by 0.32 mag at $(V-R) = 1.2$, which is the
observed mean ($V-R$) color of the galaxies in the three clusters (see
Tables 5--7). The sense of equation (2) is that the $I$ magnitudes are
brighter by 0.05 mag than the $F814W$ magnitudes of HST in the
relevant color range of the galaxies presented here.\footnote{It is
important to recall (see \S 2.2 of Sandage 1997) that the Cape/Cousins
$R$ system of magnitudes differs substantially from the Johnson $R_J$
system (Johnson 1964, 1965; Johnson et al. 1966), which is the same as
the $r_{S20}$ system introduced by Sandage \& Smith (1963). $R_J =
r_{S20}$ is also the system used by Sandage (1973, Paper V of the
redshift-distance series) for the Palomar program on the Hubble
diagram for brightest cluster galaxies. The difference between the
Cape/Cousins $R$ and $R_J$ is a function of $(V-I)$ given by equations
(25) and (26) in Sandage (1997). In the color range of interest for
early type galaxies (i.e.\ $B-V = 1.0$ to 1.1), the Johnson/Sandage
magnitudes are 0.25 mag brighter than the Cape/Cousins $R$
magnitudes. This difference must be taken into account when comparing
the aperture photometry of Postman \& Lauer (1995), which is on the
Cape/Cousins $R$ system, with that of Sandage (1973), which is on the
$R_J$ system, for cluster galaxies in common to both studies.}

\subsection{The Sample Selection}

To complete the Tolman test, we require a comparison of high-redshift
galaxies with the Postman \& Lauer (1995) sample of nearby early-type
galaxies. Therefore, we have selected our high-redshift sample by
choosing only those galaxies which are (1) morphologically classified
as early-type galaxies and (2) confirmed members of the three
clusters. To satisfy the first criteria, each galaxy had to be
morphologically classified from the high-angular-resolution HST images
as an early-type, either an elliptical or S0, galaxy by the classifiers
in Lubin et al.\ (1998, 2001).

For the second criteria, the galaxy had either to be a
spectroscopically-confirmed cluster member with an elliptical-like
(K-star absorption) spectrum, or to have the broad-band colors
characteristic of an early-type galaxy at the cluster redshift. The
latter criteria essentially uses a photometric redshift technique to
identify cluster members (for details, see Brunner \& Lubin 2000).  In
Tables 5--7, we mark those galaxies which are selected based on their
photometric, rather than their spectroscopic, redshifts. The resulting
samples contain 13, 7, and 14 galaxies in the clusters Cl 1324+3011,
Cl 1604+4304, and Cl 1604+4321, respectively.

\subsection{The Profile Data and the Three Diagnostic Functions}

The two-dimensional HST data were analyzed by the method of ellipse
fitting described in \S 2 of Paper II. Optimized ellipses (i.e.\ using
all the data around the bounding, iterated elliptical contours with
optimized pixel width) were fitted to the data by the profile-fitting
IRAF task ELLIPSE.  These fits produce intensities at a series of
increasing radii that could be quantified in several ways. The profile
intensities can be listed at (1) the semi-major axis, $a$, of each
ellipse, (2) the ``effective'' circular radius, $\sqrt{ab}$, where $b$
is the semi-minor axis, or (3) some intermediate radius such as that
at $45^{\rm o}$ from the major axis which has sometimes been used in
the literature.

In \S 4 of Paper II we showed that measuring the growth curve using
circular apertures was equivalent to using the profile data obtained
from the IRAF task ELLIPSE, integrating the data within circular areas
of an ``effective'' radius, $r = \sqrt{ab}$, and listing the data at
that $\sqrt{ab}$ radius.  Consequently, by using this ``equivalent
circular'' galaxy method, we are able to make an accurate comparison
with the local sample of Postman \& Lauer (1995) that was originally
analyzed using circular apertures.  Therefore, for all of the data
derived and listed in this paper, we have used the profile data given
by the IRAF program ELLIPSE and have integrated this profile with
circular apertures of radius, $r = \sqrt{ab}$, rather than the
semi-major axis, $a$.

The total magnitude inside this radius, which defines the growth curve
$m(r)$, follows directly from the integrations. Calculation of the
Petrosian radius $\eta$ at various $\sqrt{ab}$ radii follows from the
equivalent definition of $\eta$ of Djorgovski \& Spinrad (1981), where
$\eta$ is the difference in magnitudes between the surface brightness
averaged over a radius $r$ to the surface brightness (i.e.\ the
profile intensity) at that radius $r$ (see also Sandage \& Perelmuter
1990 for a different derivation of this equivalent definition).
     
We illustrate the calculations using the HST data for three sample
galaxies, one from each cluster studied here. Figure 1 shows the HST
image of galaxy HST \#40 in Cl 1324+3011. This is the ninth-ranked
galaxy in the list of 13 galaxies for this cluster (see Table 5). Its
total magnitude, derived from the asymptote of the growth curve and
using equation (2), is $I = 21.63$. Its half-light radius is
$\sqrt{ab} = 0\farcs{26}$.  (Note that the half-light radius for
galaxies with normal intensity profiles occurs near $\eta = 1.0$ mag;
see Sandage \& Perelmuter 1990).

The HST photometric data for this galaxy in the $F814W$ photometric
system are listed in Table 1. Column 1 gives the log of the
$\sqrt{ab}$ radius. Column 2 lists the $F814W$ surface brightness and
the rms error (in magnitudes) of the profile at that radius, as
determined by the IRAF program ELLIPSE. Column 3 gives the total
magnitude (and error) within a circular aperture of radius
$\sqrt{ab}$. This quantity was calculated by integrating the profile
data in column 2 within the ``effective'' circular aperture of radius
$\sqrt{ab}$ listed in column 1. These data define the growth curve,
$m(r)$.  Column 4 lists the mean surface brightness, $\langle SB
\rangle$, averaged over the circular aperture with radius $\sqrt{ab}$.
It is calculated by adding $5 \times {\rm log} (\sqrt{\pi ab})$ to the
magnitude in column 3. Column 5 lists the Petrosian $\eta$ value (and
error) calculated by subtracting column 4 from column 2. Column 6
lists the ellipticity, defined as $\epsilon = 1-{b\over{a}}$, of the
best-fitting ellipse at the listed radius, as determined using the
IRAF program ELLIPSE.

Figure 2 shows the four diagnostic curves for HST \#40 in Cl
1324+3011, based on the data given in Table 1. The format of these
diagrams is the same as Figures 8 and 9 of Paper II. Three of the
panels show mean surface brightness $\langle SB \rangle$, apparent
magnitude $m$, and Petrosian radius $\eta$ versus the ``effective''
circular radius, $\sqrt{ab}$ (in arcsec). The lower right panel shows
$\langle SB \rangle$ versus $\eta$.

Figure 3 shows the image of galaxy HST \#13 in Cl 1604+4304. This is
the third-ranked galaxy of the seven galaxies listed for this cluster
in Table 6. Its total magnitude, derived from the asymptote of the
growth curve and using equation (2), is $I = 20.68$. Its half-light
radius is 0\farcs{36}.  Table 2 lists the same data as Table 1 but for
HST \#13 in Cl 1604+4304. Again, the listings are given in the HST
$F814W$ photometric system. Figure 4 shows the resulting four
diagnostic diagrams.

Figure 5 shows the image of galaxy HST \#29 in Cl 1604+4321. This is
the second-ranked galaxy of the 14 galaxies listed for this cluster in
Table 7. The HST observations were made in the $F702W$ filter. The
total magnitude from the asymptote of the growth curve, the observed
Keck color of $(V-R) = 0.99$ (see Table 7), and equation (1), is $R =
22.36$. Its half-light radius is 0\farcs{35}. Table 3 lists the same
basic data as Tables 1 and 2. Figure 6 shows the four diagnostic
diagrams, with the photometry given in the $F702W$ bandpass.

These three galaxies are typical of the program galaxies, although
they are neither the brightest nor the faintest in each cluster. The
angular radii at $\eta = 1$ mag of the 34 program galaxies range from
0\farcs{57} for the largest to 0\farcs{17} for the smallest. It was
shown in Paper II that galaxies with half-light radii of 0\farcs{25}
should be free from systematic errors larger than $\Delta \langle SB
\rangle = 0.07$ mag due to the finite point spread function of WFPC2
for all $\eta \gtrsim 1.8$. Hence, the data presented here for $\eta =
1.8$ mag and larger should be free of systematic errors due to the
finite resolution of HST. For $\eta < 1.8$ mag, the data on the
smallest galaxies will not be free from some systematic error. We test
this prediction in Paper IV where the systematics of the analysis are
given for different $\eta$ values.

All 34 galaxies in the three high-redshift clusters were analyzed in
the same ways illustrated in Tables 1--3 and Figures 2, 4, \& 6. The
data measured at each designated $\eta$ value for each galaxy in our
sample, obtained by a linear interpolation in similar diagrams for
each galaxy, are listed in Tables 5--7 (see \S 4).

\section{Calculation of the $K$ Corrections for the Three 
High-Redshift Clusters}

In order to make a comparison with observations of local objects at
zero redshift, the $K$ term is a technical correction that must be
applied to any observed finite-bandwidth photometry for objects whose
spectrum is redshifted (Humason, Mayall \& Sandage 1956; Oke \&
Sandage 1968; Sandage 1995). As defined by these authors, the $K$ term
corrects the observed intensity to what would have been observed in
the rest frame of the particular photometric band defined by a given
filter with its color equation.

The observations by OPL with the Keck telescopes were made with a set
of filters whose transmission functions are close to those of the
Johnson/Cousins standard $BVRI$ system (see Figure 1 of OPL). The
filters, of course, sample the near ultraviolet spectra of the highly
redshifted cluster galaxies studied here.  For example, the $R$ and
$I$ filters with mean wavelengths of 6000\AA\ and 8000\AA,
respectively, sample the rest-frame spectral energy distributions at
the mean rest wavelengths of 3158\AA\ and 4211\AA, respectively, for a
galaxy at $z = 0.90$.

To determine what the observed intensities would have been if the
redshifted galaxies had been observed at the rest-frame wavelengths of
the Cape/Cousins $RI$ system, we must know the spectral energy
distribution (SED) of the redshifted galaxy in order to tie the
observed data at the near-UV and $B$ rest wavelengths to the standard
$R$ and $I$ rest wavelengths. Therefore, to calculate the $K$
correction for a given galaxy at redshift of e.g.\ $z = 0.90$, we need
the true SED of that galaxy over the rest wavelength range from
3000\AA\ to the limit of the transmission function of the $I$ filter
near 11000\AA\ (Figure 1 of OPL).

The $K$ corrections that have been tabulated in most of the early
literature (e.g.\ Oke \& Sandage 1968; Whitford 1971; Yoshii \&
Takahara 1988; Coleman, Wu \& Weedman 1988) used SEDs that were either
observed for local, low-redshift galaxies over a large wavelength
range or were calculated using stellar population synthesis programs
in the no-evolution case. For early-type (elliptical and S0) galaxies,
the observed SEDs of elliptical galaxies, such as NGC 4486 or NGC
4472, and/or the central bulge of Sb galaxies, such as M31, were
used. These spectra then were redshifted through the filter curves to
calculate $K_z(BVRI)$.

However, it became clear from the color indices measured by OPL and
analyzed by PLO98 and PLO01 that the observed colors of the
high-redshift galaxies used here could not be explained with
no-evolution SEDs (see Figure 7). Therefore, to calculate valid $K$
corrections for the galaxies in the three high-redshift clusters, we
require knowledge of their actual SEDs, taking into account a
luminosity evolution that is a function of wavelength (see Figures 10
\& 11 of PLO98).

We have obtained the actual SEDs through work originally performed by
Postman, Lubin \& Oke to examine the star-formation histories and the
ages of the cluster members (PLO98; PLO01). These authors compared the
observed broad-band ($BVRI$) AB magnitudes of each galaxy with a
measured redshift to synthetic SEDs calculated from the 1996 stellar
evolution code of Bruzual \& Charlot (1993) under a variety of the
star-formation scenarios.  Specifically, they used solar metallicity
($Z = 0.02$) models where the star-formation rate falls exponentially
with time. These models are referred to as tau ($\tau$) models. A
best-fit, zero-redshift SED was calculated for each galaxy using its
broad-band AB magnitudes which were blueshifted to the appropriate
rest wavelengths by a factor of $(1 + z)$ in ${\Delta
\lambda}\over{\lambda}$ (see PLO98 and PLO01 for details of the
fitting procedure).

We note that, over the observed passbands, the final best-fit spectrum
is nearly independent of the particular $\tau$ model. The largest
differences arise at rest UV wavelengths of $\lesssim 2500$\AA\ which
are most sensitive to the youngest stars. This result is illustrated
in Figure 7 where we show the best-fit spectrum from a $\tau = 0.2$
Gyr model (solid line) and a $\tau = 0.6$ Gyr model (dotted line) to
the observed AB magnitudes of galaxy HST \#11 in Cl 1324+3011. The
redshift of this galaxy is $z = 0.7580$ (PLO01).  The difference in
the $K$ correction measured from the two best-fit synthetic SEDs would
be only $\Delta K < 0.06$ mag, depending on the particular passband.
Consequently, we have adopted for this paper the results from the
$\tau = 0.2$ Gyr model of Bruzual \& Charlot (1993).

Once an individual SED was adopted for each galaxy, it was used to
calculate an individual $K$ correction at the mean redshift of the
cluster. This was done by redshifting the adopted rest-wavelength
synthetic SED by $(1 + z)$ and calculating the intensity ratio of the
redshifted SED to the non-redshifted SED after convolving each with
the appropriate filter function from Figure 1 of OPL.  Because the
effective bandwidth of the shifted SED is smaller than that for the
unshifted spectrum by $\lambda \over{1 + z}$, the non-selective
bandwidth factor of $2.5~{\rm log}(1 + z)$ mag must be added to this
intensity ratio (expressed in magnitudes). [What we have just
described is the explanation of equation (B7) in Appendix B of
Humason, Mayall \& Sandage (1956), where the theory of the $K$
correction is given in detail.]

We have calculated individual $K(R)$ and $K(I)$ values for each of the
early-type galaxies in the three clusters by using the best-fit
synthetic Bruzual \& Charlot SED based on the individual, observed AB
magnitudes. The corrections are not all identical in a given cluster
because of variations in the measured magnitudes at different
wavelengths which are caused by a combination of measuring errors due
to photon statistics and real differences in the intrinsic
SEDs. Therefore, we have averaged the individual $K$ corrections for
each cluster and have adopted the mean values listed in Table 4. The
rms variation in the individual $K$ terms, listed as $\Delta K$ in
Table 4, are small at $\Delta K \le 0.08$.

\section{$K$-Corrected Apparent Magnitudes, Mean Surface Brightnesses,
and Angular Radii for the Program Galaxies at Five $\eta$ Values}

We have measured the total magnitude $m$, mean surface brightness
$\langle SB \rangle$, and angular radius $r$ at each of the five
$\eta$ values for each galaxy in our sample. These data were derived
from the HST profile data in the way described in \S 2. However, in
order to remove some of the stochastic nature of the curves (see e.g.\
Figures 2, 4 \& 6), the measured $SB(r)$, $\langle SB(r) \rangle$,
$m(r)$, and $\eta(r)$ functions of each galaxy were smoothed with a
three-point boxcar filter.  The smoothed curves permitted the growth
curve $m(\eta)$, the mean surface brightness $\langle SB(\eta)
\rangle$, and the angular radii $r(\eta)$ functions to be interpolated
analytically at the discrete values of $\eta = \{ 1.0, 1.3, 1.5, 1.7,
2.0\}$. These values, corrected by the adopted mean $K$ terms in Table
4, are given in Tables 5--7. We have estimated the error on the total
magnitude by adding in quadrature the uncertainties associated with
the measured $K$ correction (including both the rms error on the mean
value and the uncertainty in the choice of Bruzual \& Charlot model;
see \S 3) and our ``effective circular'' galaxy approximation (see \S
4.2 of Paper II). These errors also apply to the measurements of mean
surface brightness $\langle SB \rangle$.

A reader attentive to details will notice that the entries in Tables
5--7 do not always agree precisely with

\begin{equation}
           \langle SB \rangle = m + 5~{\rm log} (\sqrt{\pi ab})
\end{equation}

\noindent which is the definition of mean surface brightness. This
equation was used, as described in \S 2.2, to calculate $\langle SB
\rangle$ from the total magnitude, $m$, and ``effective'' circular
radius, $\sqrt{ab}$, using data tables similar to Tables 1--3.
However, the entries in Tables 5--7 are obtained by interpolating the
smoothed curves of the $\langle SB \rangle$, $m$, and log $\sqrt{ab}$
as a function of $\eta$ for each of the program galaxies. In the
majority of cases, the listed log $\sqrt{ab}$ values in Tables 5--7
are within less than 0.01 dex of the requirement of equation (3).
Where there are differences, they are due to the effects of the
smoothing of the observed $\langle SB(\eta) \rangle$ and $r(\eta)$
functions which produce small deviations among the measured
quantities.  Deviations in the log of the linear radius R (used in
Paper IV) at the level of 0.007 dex, which is the rms value for the
mean deviations from equation (3) of all listed angular radii in
Tables 5--7, are totally negligible for the analysis.

The data in Tables 5--7 are the source in the search for a Tolman
signal in Paper IV. It is to be emphasized again that the data are
independent of any assumptions about cosmological parameters. They are
directly observed from the profile data. The $\eta$ values are metric
radii defined by the ratio of surface brightnesses that are measured
directly over the face of the galaxy, with no assumption as to any
value of a {\it linear} radius. The apparent magnitudes within the
designated $\eta$ radius requires only a summing of intensities in the
pixels interior to that radius.

On the other hand, use of these data for the Tolman test by comparing
the $\langle SB \rangle$ values in Tables 5--7 with those of the local
galaxies in Paper I requires knowledge of the {\it absolute}
magnitudes, $M$, and {\it linear} radii, R. These quantities are
needed to produce the diagnostic diagrams of Paper I. The calculations
of $M$ and R do require knowledge of the deceleration parameter,
$q_o$, that appears in the Mattig (1958) equations. Although the
analysis for the Tolman test is slightly degenerate because of this
requirement, we have shown in \S 5 of Paper I that the resulting
hermeneutical circularity is sufficiently small. We shall show this
again in Paper IV by comparing the Tolman signals for different
assumed $q_o$ values.

\acknowledgments 

We are grateful to Bev Oke and Marc Postman for their kindness in
permitting our use of the Oke, Postman \& Lubin photometric data for
the various calculations before the publication of the complete data
in the literature. LML was supported by NASA through Hubble Fellowship
grant HF-01095.01-97A from the Space Telescope Science Institute,
which is operated by the Association of Universities for Research in
Astronomy, Inc., under NASA contract NAS 5-26555. AS acknowledges
support for publication from NASA grants GO-5427.01-93A and
GO-06459.01-95A for work that is related to data taken with the {\it
Hubble Space Telescope}.

\newpage

\begin{deluxetable}{rccccc}
\tablewidth{0pt}
\tablenum{1}
\tablecaption{Derived Photometric Parameters for HST \#40 in Cl 1324+3011 (HST $F814W$ Band)}
\tablehead{
\colhead{Log $\sqrt{ab}$} &
\colhead{$SB \pm \Delta SB$} &
\colhead{$m \pm \Delta m$} &
\colhead{$\langle SB \rangle$} &
\colhead{$\eta \pm \Delta \eta$} &
\colhead{$\epsilon$} \\
\colhead{(arcsec)} &
\colhead{(mag/arcsec$^2$)} &
\colhead{(mag)} &
\colhead{(mag/arcsec$^2$)} &
\colhead{(mag)} &
\colhead{} \\
\colhead{(1)} &
\colhead{(2)} &
\colhead{(3)} &
\colhead{(4)} &
\colhead{(5)} &
\colhead{(6)}}
\startdata
 $ -1.003$ &  $20.05 \pm 0.01$ &  $23.070 \pm 0.014$ &  19.297  &   $0.75 \pm 0.02$ &   0.006 \\
 $ -0.710$ &  $21.15 \pm 0.01$ &  $22.503 \pm 0.010$ &  20.196  &   $0.96 \pm 0.02$ &   0.041 \\
 $ -0.560$ &  $21.89 \pm 0.01$ &  $22.266 \pm 0.008$ &  20.706  &   $1.18 \pm 0.01$ &   0.152 \\
 $ -0.448$ &  $22.42 \pm 0.01$ &  $22.108 \pm 0.007$ &  21.109  &   $1.32 \pm 0.01$ &   0.201 \\
 $ -0.355$ &  $22.93 \pm 0.01$ &  $21.993 \pm 0.007$ &  21.462  &   $1.47 \pm 0.01$ &   0.213 \\
 $ -0.284$ &  $23.29 \pm 0.02$ &  $21.913 \pm 0.006$ &  21.735  &   $1.56 \pm 0.02$ &   0.243 \\
 $ -0.222$ &  $23.63 \pm 0.02$ &  $21.847 \pm 0.006$ &  21.981  &   $1.65 \pm 0.02$ &   0.259 \\
 $ -0.163$ &  $24.00 \pm 0.02$ &  $21.793 \pm 0.006$ &  22.219  &   $1.78 \pm 0.03$ &   0.258 \\
 $ -0.106$ &  $24.47 \pm 0.04$ &  $21.750 \pm 0.006$ &  22.464  &   $2.01 \pm 0.04$ &   0.235 \\
 $ -0.058$ &  $24.80 \pm 0.05$ &  $21.718 \pm 0.006$ &  22.670  &   $2.13 \pm 0.05$ &   0.229 \\
 $ -0.017$ &  $25.17 \pm 0.07$ &  $21.694 \pm 0.006$ &  22.853  &   $2.32 \pm 0.07$ &   0.229 \\
 $  0.018$ &  $25.46 \pm 0.08$ &  $21.677 \pm 0.006$ &  23.008  &   $2.45 \pm 0.08$ &   0.240 \\
 $  0.031$ &  $25.56 \pm 0.08$ &  $21.670 \pm 0.006$ &  23.068  &   $2.49 \pm 0.08$ &   0.312 \\
 $  0.063$ &  $25.80 \pm 0.10$ &  $21.656 \pm 0.006$ &  23.215  &   $2.59 \pm 0.10$ &   0.312 \\
 $  0.101$ &  $26.05 \pm 0.10$ &  $21.640 \pm 0.006$ &  23.389  &   $2.66 \pm 0.10$ &   0.286 \\
 $  0.129$ &  $26.30 \pm 0.14$ &  $21.629 \pm 0.006$ &  23.519  &   $2.78 \pm 0.14$ &   0.286 \\
 $  0.156$ &  $26.83 \pm 0.16$ &  $21.623 \pm 0.006$ &  23.643  &   $3.19 \pm 0.16$ &   0.286 \\
 $  0.180$ &  $26.80 \pm 0.20$ &  $21.615 \pm 0.006$ &  23.760  &   $3.04 \pm 0.20$ &   0.286 \\
 $  0.204$ &  $27.34 \pm 0.36$ &  $21.611 \pm 0.006$ &  23.873  &   $3.46 \pm 0.36$ &   0.286 \\
\enddata
\end{deluxetable}

\begin{deluxetable}{rccccc}
\tablewidth{0pt}
\tablenum{2}
\tablecaption{Derived Photometric Parameters for HST \#13 in Cl 1604+4304 (HST $F814W$ Band)}
\tablehead{
\colhead{Log $\sqrt{ab}$} &
\colhead{$SB \pm \Delta SB$} &
\colhead{$m \pm \Delta m$} &
\colhead{$\langle SB \rangle$} &
\colhead{$\eta \pm \Delta \eta$} &
\colhead{$\epsilon$} \\
\colhead{(arcsec)} &
\colhead{(mag/arcsec$^2$)} &
\colhead{(mag)} &
\colhead{(mag/arcsec$^2$)} &
\colhead{(mag)} &
\colhead{} \\
\colhead{(1)} &
\colhead{(2)} &
\colhead{(3)} &
\colhead{(4)} &
\colhead{(5)} &
\colhead{(6)}}
\startdata
 $-1.029$ &  $20.07 \pm 0.01$ &  $23.223 \pm 0.006$ &  19.319  &   $0.75 \pm 0.01$ &   0.119 \\
 $-0.734$ &  $20.59 \pm 0.01$ &  $22.369 \pm 0.006$ &  19.942  &   $0.65 \pm 0.01$ &   0.142 \\
 $-0.552$ &  $21.18 \pm 0.01$ &  $21.923 \pm 0.005$ &  20.407  &   $0.77 \pm 0.01$ &   0.117 \\
 $-0.425$ &  $21.73 \pm 0.01$ &  $21.666 \pm 0.004$ &  20.783  &   $0.95 \pm 0.01$ &   0.111 \\
 $-0.326$ &  $22.24 \pm 0.01$ &  $21.498 \pm 0.004$ &  21.109  &   $1.13 \pm 0.01$ &   0.103 \\
 $-0.246$ &  $22.71 \pm 0.01$ &  $21.384 \pm 0.003$ &  21.395  &   $1.32 \pm 0.01$ &   0.100 \\
 $-0.180$ &  $23.15 \pm 0.01$ &  $21.305 \pm 0.003$ &  21.645  &   $1.50 \pm 0.01$ &   0.104 \\
 $-0.122$ &  $23.54 \pm 0.01$ &  $21.244 \pm 0.003$ &  21.875  &   $1.66 \pm 0.01$ &   0.103 \\
 $-0.077$ &  $23.81 \pm 0.01$ &  $21.201 \pm 0.003$ &  22.056  &   $1.76 \pm 0.01$ &   0.129 \\
 $-0.031$ &  $24.17 \pm 0.02$ &  $21.162 \pm 0.003$ &  22.252  &   $1.92 \pm 0.02$ &   0.125 \\
 $ 0.004$ &  $24.41 \pm 0.02$ &  $21.136 \pm 0.003$ &  22.398  &   $2.01 \pm 0.02$ &   0.153 \\
 $ 0.038$ &  $24.65 \pm 0.02$ &  $21.112 \pm 0.003$ &  22.546  &   $2.11 \pm 0.02$ &   0.165 \\
 $ 0.075$ &  $24.90 \pm 0.02$ &  $21.088 \pm 0.003$ &  22.707  &   $2.19 \pm 0.02$ &   0.156 \\
 $ 0.102$ &  $25.12 \pm 0.03$ &  $21.073 \pm 0.003$ &  22.824  &   $2.30 \pm 0.03$ &   0.178 \\
 $ 0.132$ &  $25.44 \pm 0.04$ &  $21.058 \pm 0.003$ &  22.959  &   $2.48 \pm 0.04$ &   0.178 \\
 $ 0.185$ &  $25.80 \pm 0.06$ &  $21.035 \pm 0.003$ &  23.201  &   $2.60 \pm 0.06$ &   0.079 \\
 $ 0.207$ &  $25.92 \pm 0.07$ &  $21.025 \pm 0.003$ &  23.303  &   $2.61 \pm 0.07$ &   0.095 \\
 $ 0.218$ &  $25.89 \pm 0.06$ &  $21.020 \pm 0.003$ &  23.353  &   $2.54 \pm 0.06$ &   0.150 \\
\enddata
\end{deluxetable}

\begin{deluxetable}{rccccc}
\tablewidth{0pt}
\tablenum{3}
\tablecaption{Derived Photometric Parameters for HST \#29 in Cl 1604+4321 (HST $F702W$ Band)}
\tablehead{
\colhead{Log $\sqrt{ab}$} &
\colhead{$SB \pm \Delta SB$} &
\colhead{$m \pm \Delta m$} &
\colhead{$\langle SB \rangle$} &
\colhead{$\eta \pm \Delta \eta$} &
\colhead{$\epsilon$} \\
\colhead{(arcsec)} &
\colhead{(mag/arcsec$^2$)} &
\colhead{(mag)} &
\colhead{(mag/arcsec$^2$)} &
\colhead{(mag)} &
\colhead{} \\
\colhead{(1)} &
\colhead{(2)} &
\colhead{(3)} &
\colhead{(4)} &
\colhead{(5)} &
\colhead{(6)}}
\startdata
 $-1.051$  & $21.27 \pm 0.01$ &  $24.536 \pm 0.006$ &  20.522  &   $0.75 \pm 0.01$ &   0.204 \\
 $-0.715$  & $21.64 \pm 0.01$ &  $23.418 \pm 0.004$ &  21.083  &   $0.56 \pm 0.01$ &   0.066 \\
 $-0.539$  & $22.08 \pm 0.01$ &  $22.907 \pm 0.004$ &  21.453  &   $0.62 \pm 0.01$ &   0.066 \\
 $-0.414$  & $22.64 \pm 0.01$ &  $22.623 \pm 0.003$ &  21.795  &   $0.84 \pm 0.01$ &   0.064 \\
 $-0.319$  & $23.21 \pm 0.01$ &  $22.456 \pm 0.003$ &  22.106  &   $1.11 \pm 0.01$ &   0.070 \\
 $-0.240$  & $23.74 \pm 0.01$ &  $22.347 \pm 0.003$ &  22.392  &   $1.35 \pm 0.02$ &   0.071 \\
 $-0.176$  & $24.19 \pm 0.03$ &  $22.274 \pm 0.004$ &  22.637  &   $1.56 \pm 0.03$ &   0.085 \\
 $-0.120$  & $24.53 \pm 0.03$ &  $22.216 \pm 0.004$ &  22.859  &   $1.67 \pm 0.03$ &   0.094 \\
 $-0.072$  & $24.94 \pm 0.04$ &  $22.175 \pm 0.004$ &  23.057  &   $1.88 \pm 0.04$ &   0.108 \\
 $-0.046$  & $25.04 \pm 0.06$ &  $22.152 \pm 0.004$ &  23.166  &   $1.88 \pm 0.06$ &   0.183 \\
 $ 0.008$  & $25.26 \pm 0.06$ &  $22.106 \pm 0.004$ &  23.388  &   $1.88 \pm 0.06$ &   0.137 \\
 $ 0.046$  & $25.64 \pm 0.07$ &  $22.080 \pm 0.005$ &  23.550  &   $2.09 \pm 0.07$ &   0.137 \\
 $ 0.067$  & $25.80 \pm 0.08$ &  $22.065 \pm 0.005$ &  23.644  &   $2.16 \pm 0.08$ &   0.187 \\
 $ 0.099$  & $26.08 \pm 0.11$ &  $22.046 \pm 0.005$ &  23.786  &   $2.29 \pm 0.11$ &   0.187 \\
 $ 0.129$  & $26.00 \pm 0.09$ &  $22.025 \pm 0.005$ &  23.915  &   $2.09 \pm 0.09$ &   0.187 \\
\enddata
\end{deluxetable}

\begin{deluxetable}{rccc}
\tablewidth{0pt}
\tablenum{4}
\tablecaption{Adopted $K$-Corrections in Cape/Cousins $R$ and $I$ 
Bands for the Three Clusters}
\tablehead{
\colhead{Cluster} &
\colhead{$z$} &
\colhead{$K(R) \pm \Delta K(R)$} &
\colhead{$K(I) \pm \Delta K(I)$} \\
\colhead{} &
\colhead{} &
\colhead{(mag)} &
\colhead{(mag)}}
\startdata
1324+3011 & 0.7565 & $1.59 \pm 0.04$ & $0.71 \pm 0.03$ \\
1604+4304 & 0.8967 & $1.70 \pm 0.08$ & $0.80 \pm 0.07$ \\
1604+4321 & 0.9243 & $1.89 \pm 0.06$ & $1.00 \pm 0.05$ \\
\enddata
\end{deluxetable}

\newpage
\voffset +1.00in

\begin{deluxetable}{cccccccccccccccccc}
\rotate
\scriptsize
\tablewidth{0pt}
\tabletypesize{\scriptsize}
\tablenum{5}
\tablecaption{Observed Data for Cl 1324+3011 [$I$ Band; $K(I) = 0.71$]}
\tablehead{
\colhead{Gal} &
\multicolumn{3}{c}{$\eta = 1.0$} &
\multicolumn{3}{c}{$\eta = 1.3$} &
\multicolumn{3}{c}{$\eta = 1.5$} &
\multicolumn{3}{c}{$\eta = 1.7$} &
\multicolumn{3}{c}{$\eta = 2.0$} &
\multicolumn{2}{c}{Colors\tablenotemark{a}} \\
\colhead{\#} &
\colhead{$I$\tablenotemark{b}} &
\colhead{$\langle SB \rangle$\tablenotemark{b}} &
\colhead{Log $r''$} &
\colhead{$I$\tablenotemark{b}} &
\colhead{$\langle SB \rangle$\tablenotemark{b}} &
\colhead{Log $r''$} &
\colhead{$I$\tablenotemark{b}} &
\colhead{$\langle SB \rangle$\tablenotemark{b}} &
\colhead{Log $r''$} &
\colhead{$I$\tablenotemark{b}} &
\colhead{$\langle SB \rangle$\tablenotemark{b}} &
\colhead{Log $r''$} &
\colhead{$I$\tablenotemark{b}} &
\colhead{$\langle SB \rangle$\tablenotemark{b}} &
\colhead{Log $r''$} &
\colhead{$V-R$} &
\colhead{$R-I$}}
\startdata
9  & $20.34\pm 0.07$ & $19.11$ & $-0.485$ & $19.54\pm 0.05$ & $20.75$ & $-0.005$ & $19.11 \pm 0.04$ & $21.87$ & $+0.384$ & \nodata         & \nodata & \nodata  & \nodata          & \nodata & \nodata  & $1.16$ & $1.43$ \\
11 & $20.36\pm 0.07$ & $19.89$ & $-0.336$ & $19.70\pm 0.05$ & $21.21$ & $+0.054$ & $19.46 \pm 0.04$ & $21.82$ & $+0.226$ & $19.39\pm 0.04$ & $22.04$ & $+0.312$ & \nodata          & \nodata & \nodata  & $1.21$ & $1.42$ \\
12 & $20.26\pm 0.06$ & $20.27$ & $-0.243$ & $20.03\pm 0.05$ & $20.73$ & $-0.108$ & $19.94 \pm 0.05$ & $20.97$ & $-0.044$ & $19.89\pm 0.04$ & $21.15$ & $+0.005$ & $19.84 \pm 0.04$ & $21.39$ & $+0.058$ & $1.15$ & $1.44$ \\
18 & $20.88\pm 0.07$ & $19.65$ & $-0.486$ & $20.52\pm 0.05$ & $20.39$ & $-0.272$ & $20.37 \pm 0.05$ & $20.81$ & $-0.161$ & $20.25\pm 0.05$ & $21.24$ & $-0.051$ & $20.11 \pm 0.04$ & $21.95$ & $+0.119$ & $1.17$ & $1.40$ \\
21 & $21.12\pm 0.08$ & $19.36$ & $-0.583$ & $20.66\pm 0.06$ & $20.31$ & $-0.307$ & $20.49 \pm 0.05$ & $20.81$ & $-0.176$ & $20.37\pm 0.05$ & $21.26$ & $-0.067$ & $20.28 \pm 0.04$ & $21.69$ & $+0.042$ & $1.21$ & $1.43$ \\
26 & $21.33\pm 0.07$ & $20.20$ & $-0.464$ & $20.69\pm 0.05$ & $21.44$ & $-0.094$ & $20.55 \pm 0.05$ & $21.82$ & $+0.002$ & $20.45\pm 0.05$ & $22.16$ & $+0.085$ & $20.40 \pm 0.04$ & $22.41$ & $+0.160$ & $1.14$ & $1.44$ \\
29 & $21.10\pm 0.06$ & $20.36$ & $-0.391$ & $20.74\pm 0.05$ & $21.08$ & $-0.168$ & $20.57 \pm 0.04$ & $21.55$ & $-0.059$ & $20.32\pm 0.04$ & $22.45$ & $+0.177$ & $20.25 \pm 0.04$ & $22.81$ & $+0.264$ & $1.16$ & $1.45$ \\
30\tablenotemark{c}& $21.15\pm 0.07$ & $20.52$ & $-0.364$ & $20.96\pm 0.05$ & $20.93$ & $-0.253$ & $20.85 \pm 0.05$ & $21.22$ & $-0.170$ & $20.75\pm 0.04$ & $21.62$ & $-0.069$ & $20.68 \pm 0.04$ & $21.95$ & $+0.008$ & $1.13$ & $1.41$ \\
40 & $21.79\pm 0.07$ & $19.42$ & $-0.681$ & $21.38\pm 0.05$ & $20.28$ & $-0.463$ & $21.20 \pm 0.05$ & $20.81$ & $-0.322$ & $21.08\pm 0.04$ & $21.28$ & $-0.208$ & $20.99 \pm 0.04$ & $21.72$ & $-0.099$ & $0.99$ & $1.38$ \\
55\tablenotemark{c} & $22.14\pm 0.07$ & $20.50$ & $-0.562$ & $21.88\pm 0.05$ & $21.04$ & $-0.408$ & $21.79 \pm 0.05$ & $21.31$ & $-0.339$ & $21.74\pm 0.04$ & $21.52$ & $-0.291$ & $21.66 \pm 0.03$ & $21.94$ & $-0.200$ & $1.16$ & $1.34$ \\
59 & $21.97\pm 0.08$ & $19.73$ & $-0.665$ & $21.68\pm 0.05$ & $20.30$ & $-0.515$ & $21.58 \pm 0.04$ & $20.57$ & $-0.445$ & $21.53\pm 0.04$ & $20.81$ & $-0.384$ & $21.46 \pm 0.04$ & $21.16$ & $-0.305$ & $1.19$ & $1.34$ \\
69 & $22.09\pm 0.04$ & $19.24$ & $-0.801$ & $21.86\pm 0.06$ & $19.76$ & $-0.654$ & $21.77 \pm 0.04$ & $20.05$ & $-0.583$ & $21.71\pm 0.04$ & $20.29$ & $-0.528$ & $21.65 \pm 0.03$ & $20.61$ & $-0.455$ & $0.99$ & $1.41$ \\
74\tablenotemark{c} & $22.30\pm 0.07$ & $21.28$ & $-0.440$ & $22.00\pm 0.05$ & $21.91$ & $-0.262$ & $21.85 \pm 0.05$ & $22.34$ & $-0.146$ & $21.65\pm 0.04$ & $23.03$ & $+0.028$ & $21.58 \pm 0.04$ & $23.35$ & $+0.106$ & $0.95$ & $1.48$ \\
\enddata
\tablenotetext{a}{Keck colors computed from the total magnitudes measured in matched apertures for each individual band (see \S2.2 of Brunner \& Lubin 2000).}
\tablenotetext{b}{Values have been corrected with the appropriate $K$ correction (see Table 4).}
\tablenotetext{c}{Member galaxy selected based on its photometric redshift (see Brunner \& Lubin 2000).}

\end{deluxetable}

\newpage
\voffset +1.00in

\begin{deluxetable}{cccccccccccccccccc}
\rotate
\scriptsize
\tablewidth{0pt}
\tabletypesize{\scriptsize}
\tablenum{6}
\tablecaption{Observed Data for Cl 1604+4304 [$I$ Band; $K(I) = 0.80$]}
\tablehead{
\colhead{Gal} &
\multicolumn{3}{c}{$\eta = 1.0$} &
\multicolumn{3}{c}{$\eta = 1.3$} &
\multicolumn{3}{c}{$\eta = 1.5$} &
\multicolumn{3}{c}{$\eta = 1.7$} &
\multicolumn{3}{c}{$\eta = 2.0$} &
\multicolumn{2}{c}{Colors\tablenotemark{a}} \\
\colhead{\#} &
\colhead{$I$\tablenotemark{b}} &
\colhead{$\langle SB \rangle$\tablenotemark{b}} &
\colhead{Log $r''$} &
\colhead{$I$\tablenotemark{b}} &
\colhead{$\langle SB \rangle$\tablenotemark{b}} &
\colhead{Log $r''$} &
\colhead{$I$\tablenotemark{b}} &
\colhead{$\langle SB \rangle$\tablenotemark{b}} &
\colhead{Log $r''$} &
\colhead{$I$\tablenotemark{b}} &
\colhead{$\langle SB \rangle$\tablenotemark{b}} &
\colhead{Log $r''$} &
\colhead{$I$\tablenotemark{b}} &
\colhead{$\langle SB \rangle$\tablenotemark{b}} &
\colhead{Log $r''$} &
\colhead{$V-R$} &
\colhead{$R-I$}}
\startdata
9  & $20.73\pm 0.10$ & $20.59$ & $-0.273$ & $20.25\pm 0.10$ & $21.55$ & $+0.014$ & $20.08\pm 0.09$ & $22.03$ & $+0.138$ & $19.99\pm 0.09$ & $22.36$ & $+0.225$ & $19.93\pm 0.09$ & $22.59$ & $+0.280$ & $1.31$ & $1.61$ \\
10\tablenotemark{c} & $20.85\pm 0.10$ & $20.90$ & $-0.237$ & $20.38\pm 0.10$ & $21.82$ & $+0.039$ & $20.26\pm 0.09$ & $22.16$ & $+0.134$ & $20.17\pm 0.09$ & $22.46$ & $+0.207$ & \nodata         & \nodata & \nodata  & $1.23$ & $1.72$ \\
13 & $20.80\pm 0.10$ & $20.00$ & $-0.399$ & $20.56\pm 0.10$ & $20.50$ & $-0.255$ & $20.46\pm 0.09$ & $20.80$ & $-0.178$ & $20.38\pm 0.09$ & $21.10$ & $-0.103$ & $20.29\pm 0.09$ & $21.53$ & $+0.000$ & $1.26$ & $1.63$ \\
34\tablenotemark{c} & $21.91\pm 0.10$ & $20.98$ & $-0.429$ & $21.62\pm 0.09$ & $21.60$ & $-0.250$ & $21.47\pm 0.09$ & $22.03$ & $-0.139$ & $21.39\pm 0.09$ & $22.32$ & $-0.058$ & $21.33\pm 0.09$ & $22.60$ & $+0.006$ & $1.19$ & $1.71$ \\
46\tablenotemark{c} & $21.96\pm 0.11$ & $20.40$ & $-0.546$ & $21.72\pm 0.10$ & $20.90$ & $-0.411$ & $21.63\pm 0.09$ & $21.17$ & $-0.335$ & $21.55\pm 0.09$ & $21.46$ & $-0.263$ & $21.47\pm 0.09$ & $21.90$ & $-0.163$ & $1.15$ & $1.71$ \\  
50 & $22.14\pm 0.11$ & $21.11$ & $-0.445$ & $21.79\pm 0.10$ & $21.83$ & $-0.235$ & $21.58\pm 0.09$ & $22.43$ & $-0.069$ & $21.50\pm 0.09$ & $22.73$ & $-0.005$ & $21.37\pm 0.09$ & $23.38$ & $+0.153$ & $1.26$ & $1.68$ \\
58\tablenotemark{c} & $22.25\pm 0.10$ & $20.65$ & $-0.553$ & $22.03\pm 0.09$ & $21.11$ & $-0.425$ & $21.94\pm 0.09$ & $21.37$ & $-0.355$ & $21.87\pm 0.09$ & $21.63$ & $-0.292$ & $21.79\pm 0.09$ & $22.03$ & $-0.199$ & $1.15$ & $1.74$ \\
\enddata
\tablenotetext{a}{Keck colors computed from the total magnitudes measured in matched apertures for each individual band (see \S2.2 of Brunner \& Lubin 2000).}
\tablenotetext{b}{Values have been corrected with the appropriate $K$ correction (see Table 4).}
\tablenotetext{c}{Member galaxy selected based on its photometric redshift (see Brunner \& Lubin 2000).}

\end{deluxetable}
\newpage
\voffset +1.00in

\begin{deluxetable}{cccccccccccccccccc}
\rotate
\scriptsize
\tablewidth{0pt}
\tabletypesize{\scriptsize}
\tablenum{7}
\tablecaption{Observed Data for Cl 1604+4321 [$R$ Band; $K(R) = 1.89$]}
\tablehead{
\colhead{Gal} &
\multicolumn{3}{c}{$\eta = 1.0$} &
\multicolumn{3}{c}{$\eta = 1.3$} &
\multicolumn{3}{c}{$\eta = 1.5$} &
\multicolumn{3}{c}{$\eta = 1.7$} &
\multicolumn{3}{c}{$\eta = 2.0$} &
\multicolumn{2}{c}{Colors\tablenotemark{a}} \\
\colhead{\#} &
\colhead{$R$\tablenotemark{b}} &
\colhead{$\langle SB \rangle$\tablenotemark{b}} &
\colhead{Log $r''$} &
\colhead{$R$\tablenotemark{b}} &
\colhead{$\langle SB \rangle$\tablenotemark{b}} &
\colhead{Log $r''$} &
\colhead{$R$\tablenotemark{b}} &
\colhead{$\langle SB \rangle$\tablenotemark{b}} &
\colhead{Log $r''$} &
\colhead{$R$\tablenotemark{b}} &
\colhead{$\langle SB \rangle$\tablenotemark{b}} &
\colhead{Log $r''$} &
\colhead{$R$\tablenotemark{b}} &
\colhead{$\langle SB \rangle$\tablenotemark{b}} &
\colhead{Log $r''$} &
\colhead{$V-R$} &
\colhead{$R-I$}}
\startdata
23\tablenotemark{c} & $20.91\pm 0.10$ & $18.48$ & $-0.702$ & $20.58\pm 0.09$ & $19.16$ & $-0.517$ & $20.47\pm 0.09$ & $19.49$ & $-0.433$ & $20.39\pm 0.08$ & $19.81$ & $-0.355$ & $20.26\pm 0.08$ & $20.48$ & $-0.197$ &	   1.13	& 1.51\\
29\tablenotemark{c} & $20.93\pm 0.10$ & $20.35$ & $-0.358$ & $20.76\pm 0.09$ & $20.71$ & $-0.252$ & $20.67\pm 0.09$ & $20.97$ & $-0.184$ & $20.60\pm 0.08$ & $21.23$ & $-0.121$ & $20.48\pm 0.08$ & $21.84$ & $+0.029$ &	   0.99	& 1.56\\
36 & $21.58\pm 0.10$ & $20.24$ & $-0.506$ & $21.22\pm 0.09$ & $20.97$ & $-0.294$ & $21.10\pm 0.08$ & $21.34$ & $-0.199$ & $20.99\pm 0.08$ & $21.74$ & $-0.117$ & $20.89\pm 0.08$ & $22.28$ & $+0.040$ &	   1.25	& 1.66\\
38 & $21.41\pm 0.10$ & $20.21$ & $-0.475$ & $21.17\pm 0.09$ & $20.71$ & $-0.343$ & $21.07\pm 0.08$ & $21.02$ & $-0.260$ & $20.98\pm 0.08$ & $21.39$ & $-0.159$ & $20.84\pm 0.08$ & $22.02$ & $-0.008$ &	   1.06	& 1.56\\
44\tablenotemark{c} & $21.57\pm 0.09$ & $21.06$ & $-0.340$ & $21.20\pm 0.09$ & $21.76$ & $-0.136$ & $21.09\pm 0.09$ & $22.10$ & $-0.045$ & $21.00\pm 0.08$ & $22.40$ & $+0.033$ & $20.92\pm 0.08$ & $22.82$ & $+0.132$ &	   1.20	& 1.74\\
46 & $21.74\pm 0.10$ & $21.20$ & $-0.345$ & $21.41\pm 0.09$ & $21.89$ & $-0.148$ & $21.27\pm 0.08$ & $22.27$ & $-0.047$ & $21.18\pm 0.08$ & $22.62$ & $+0.040$ & \nodata         & \nodata & \nodata  &	   0.98	& 1.60\\
49 & $22.09\pm 0.10$ & $20.07$ & $-0.623$ & $21.48\pm 0.09$ & $21.33$ & $-0.279$ & $21.33\pm 0.09$ & $21.77$ & $-0.156$ & $21.24\pm 0.08$ & $22.10$ & $-0.073$ & $21.16\pm 0.08$ & $22.46$ & $+0.004$ &	   1.34	& 1.70\\
57 & $21.89\pm 0.10$ & $21.65$ & $-0.290$ & $21.46\pm 0.09$ & $22.50$ & $-0.039$ & $21.25\pm 0.09$ & $23.05$ & $+0.103$ & $21.16\pm 0.08$ & $23.39$ & $+0.206$ & $21.12\pm 0.08$ & $23.58$ & $+0.245$ &	   1.29	& 1.58\\
58 & $22.16\pm 0.10$ & $20.38$ & $-0.581$ & $21.80\pm 0.09$ & $21.11$ & $-0.379$ & $21.59\pm 0.09$ & $21.72$ & $-0.219$ & $21.37\pm 0.08$ & $22.54$ & $-0.013$ & $21.22\pm 0.08$ & $23.25$ & $+0.157$ &	   1.24	& 1.61\\
65 & $21.94\pm 0.10$ & $19.83$ & $-0.648$ & $21.64\pm 0.09$ & $20.44$ & $-0.485$ & $21.54\pm 0.08$ & $20.74$ & $-0.411$ & $21.47\pm 0.08$ & $21.01$ & $-0.336$ & $21.40\pm 0.08$ & $21.43$ & $-0.259$ &	   1.09	& 1.54\\
78\tablenotemark{c} & $22.51\pm 0.10$ & $20.70$ & $-0.590$ & $22.18\pm 0.09$ & $21.40$ & $-0.397$ & $22.04\pm 0.09$ & $21.80$ & $-0.283$ & $21.90\pm 0.08$ & $22.30$ & $-0.167$ & $21.87\pm 0.08$ & $22.48$ & $-0.125$ &	   1.21	& 1.70\\
112\tablenotemark{c}& $22.75\pm 0.10$ & $20.27$ & $-0.707$ & $22.44\pm 0.09$ & $20.94$ & $-0.543$ & $22.34\pm 0.08$ & $21.27$ & $-0.449$ & $22.25\pm 0.08$ & $21.62$ & $-0.358$ & $22.15\pm 0.08$ & $22.12$ & $-0.259$ &	   1.21	& 1.66\\
115\tablenotemark{c}& $22.88\pm 0.09$ & $21.09$ & $-0.601$ & $22.49\pm 0.10$ & $21.87$ & $-0.370$ & $22.40\pm 0.08$ & $22.14$ & $-0.308$ & $22.33\pm 0.08$ & $22.44$ & $-0.208$ & $22.20\pm 0.08$ & $22.95$ & $-0.066$ &	   1.25	& 1.67\\
144\tablenotemark{c}& $23.03\pm 0.10$ & $20.74$ & $-0.680$ & $22.67\pm 0.09$ & $21.49$ & $-0.465$ & $22.56\pm 0.09$ & $21.82$ & $-0.390$ & $22.46\pm 0.08$ & $22.21$ & $-0.294$ & $22.35\pm 0.08$ & $22.78$ & $-0.159$ &	   1.19	& 1.71\\
\enddata
\tablenotetext{a}{Observed Keck colors computed from the total magnitudes measured in matched apertures for each individual band (see \S2.2 of Brunner \& Lubin 2000).}
\tablenotetext{b}{Values have been corrected with the appropriate $K$ correction (see Table 4).}
\tablenotetext{c}{Member galaxy selected based on its photometric redshift (see Brunner \& Lubin 2000).}

\end{deluxetable}

\newpage

\begin{figure}
\plotone{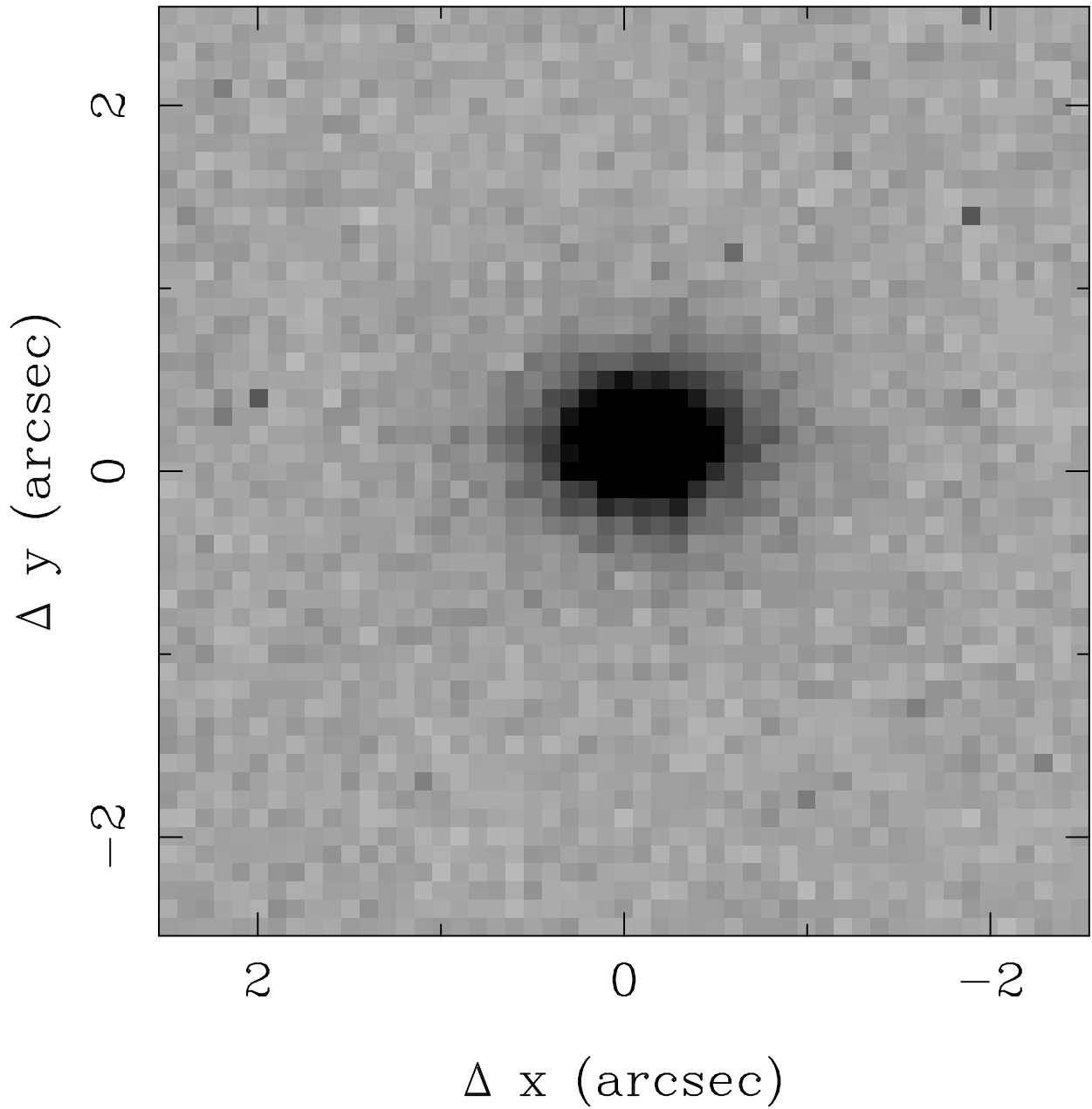}
\caption{HST image of the galaxy HST \#40 in Cl 1324 + 3011 ($z =
0.76$) in the $F814W$ bandpass. The half-light radius is $\sqrt{ab} =
0\farcs{26}$. The box size is $5'' \times 5''$.}
\end{figure}

\begin{figure}
\plotone{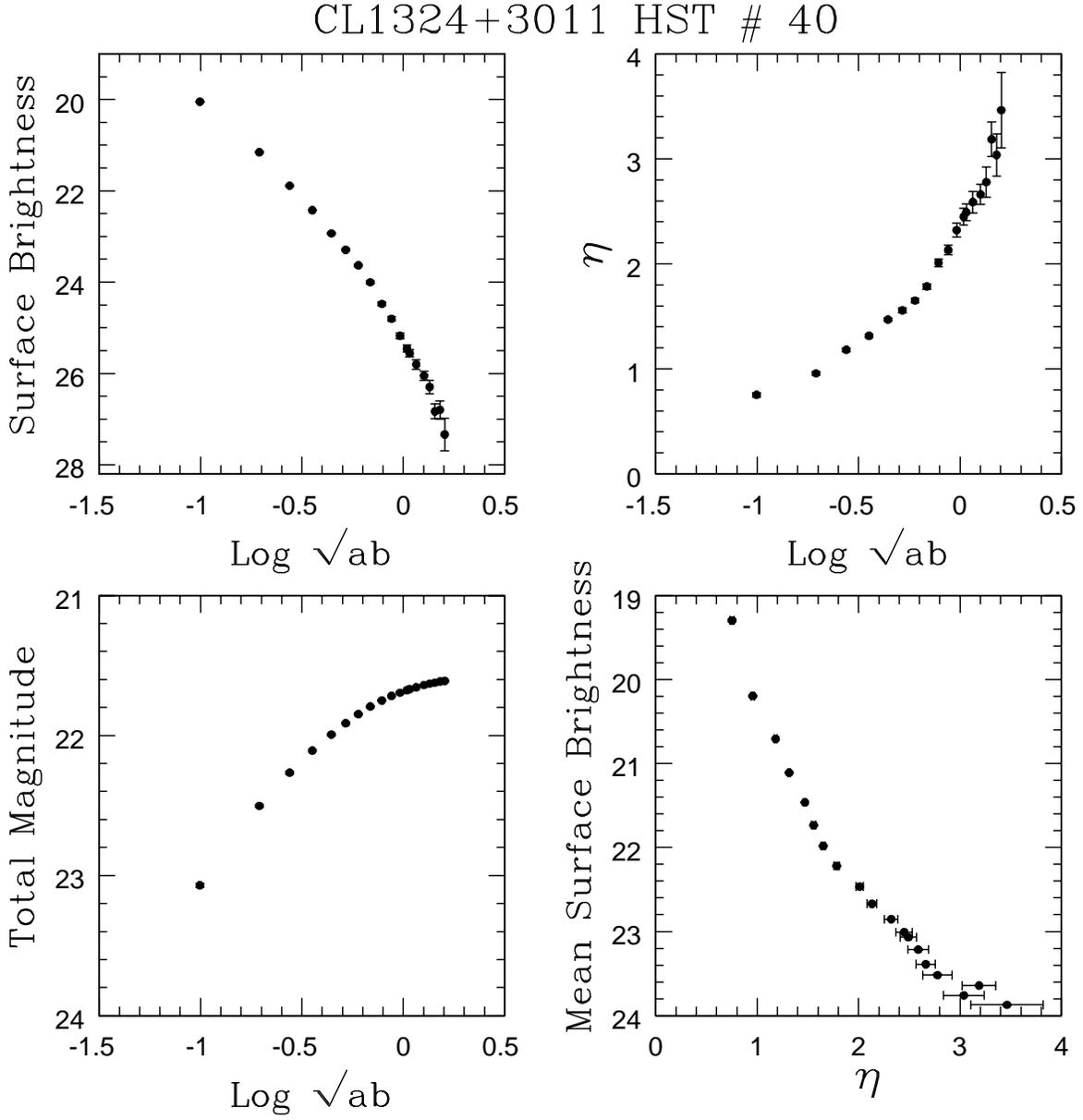}
\caption{The four diagnostic diagrams for galaxy HST \#40 in Cl
1324+3011 from the data given in Table 1. The photometry is measured
in the $F814W$ bandpass of the HST filter system. The relation between
$F814W$ and Cousins $I$ magnitudes is given in equation (2). The units
of the surface brightness are magnitudes per square arcseconds, and
the units of $\sqrt{ab}$ are in arcseconds.}
\end{figure}

\begin{figure}
\plotone{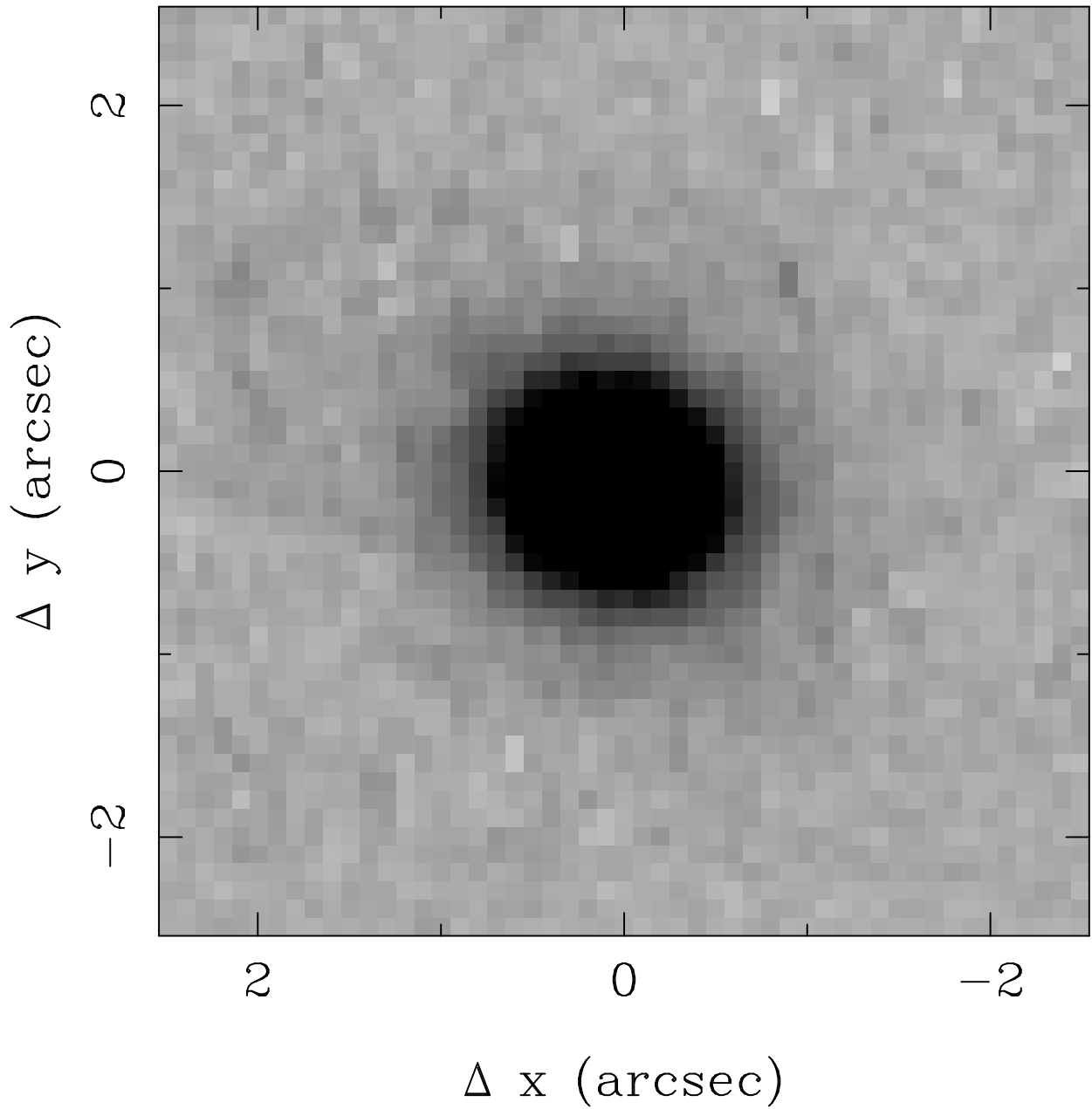}
\caption{HST image of the galaxy HST \#13 in Cl 1604+4304 ($z = 0.90$)
in the $F814$ bandpass. The half-light radius is $\sqrt{ab} =
0\farcs{36}$. The box size is $5'' \times 5''$.}
\end{figure}

\begin{figure}
\plotone{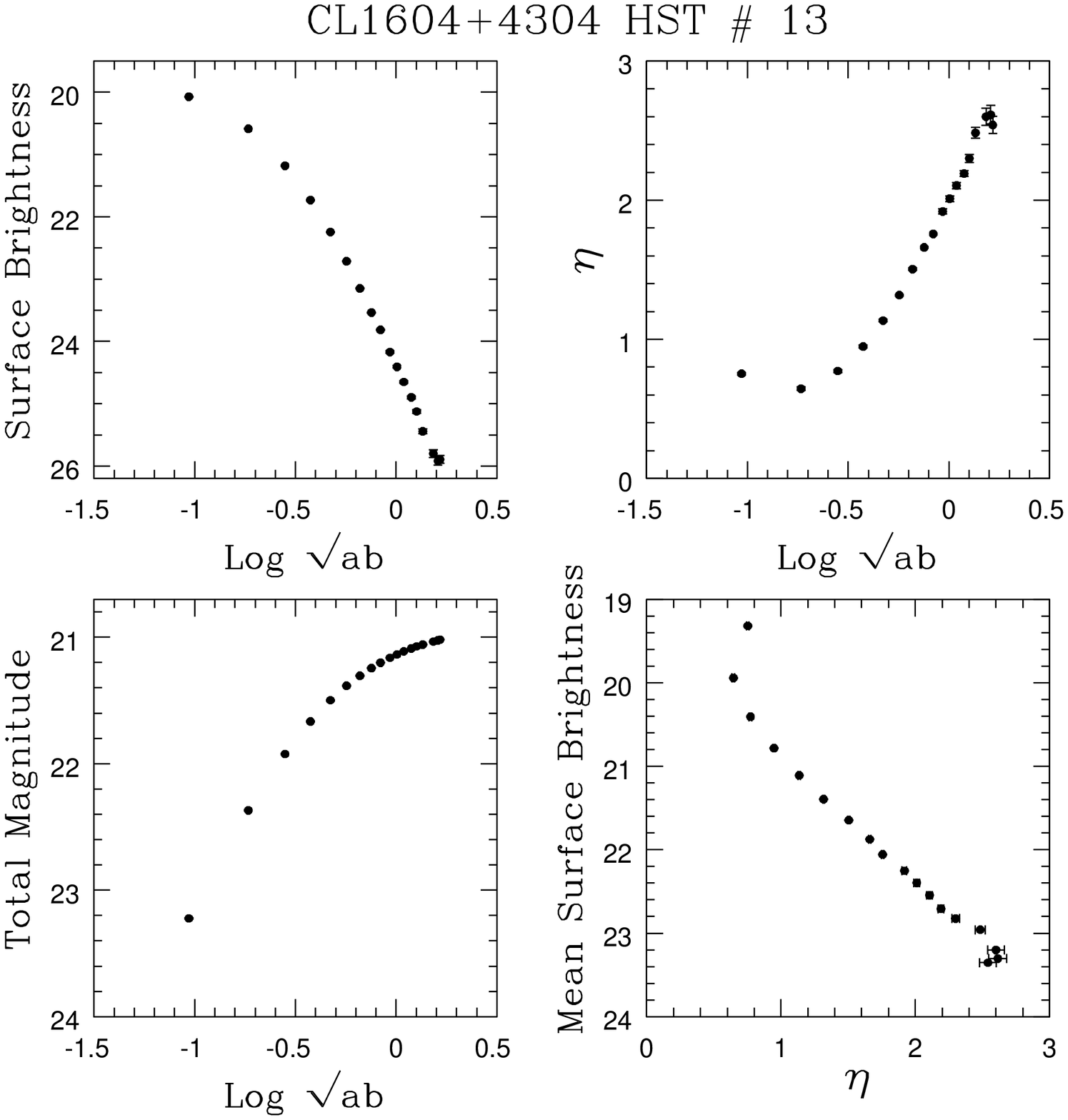}
\caption{Same as Figure 2 but for galaxy HST \#13 in Cl 1604+4304.
The photometry is given in the $F814W$ filter system.}
\end{figure}

\begin{figure}
\plotone{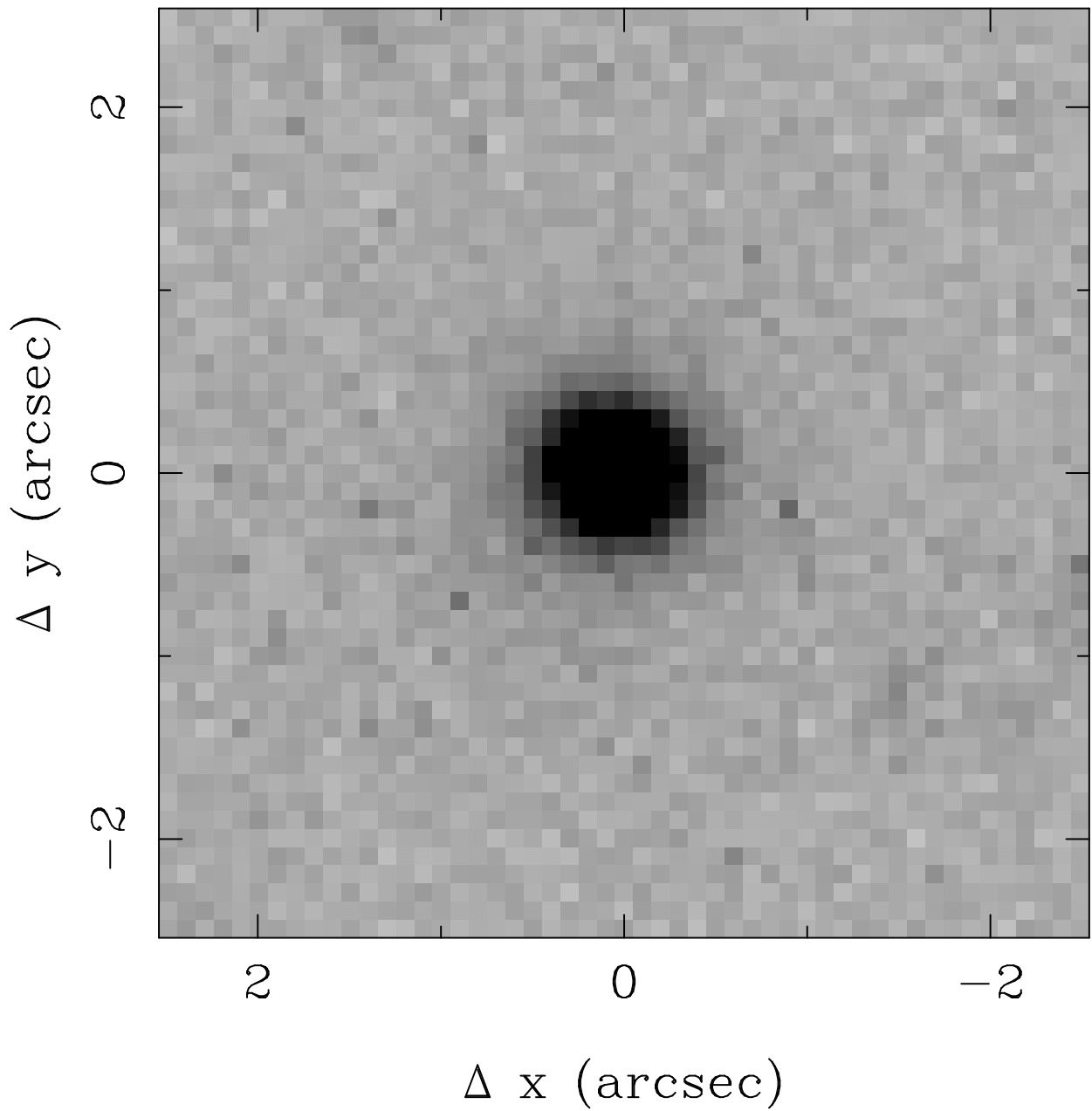}
\caption{HST image of the galaxy HST \#29 in Cl 1604+4321 ($z = 0.92$)
in the $F702W$ bandpass. The half-light radius is $\sqrt{ab} =
0\farcs{35}$. The box size is $5'' \times 5''$.}
\end{figure}

\begin{figure}
\plotone{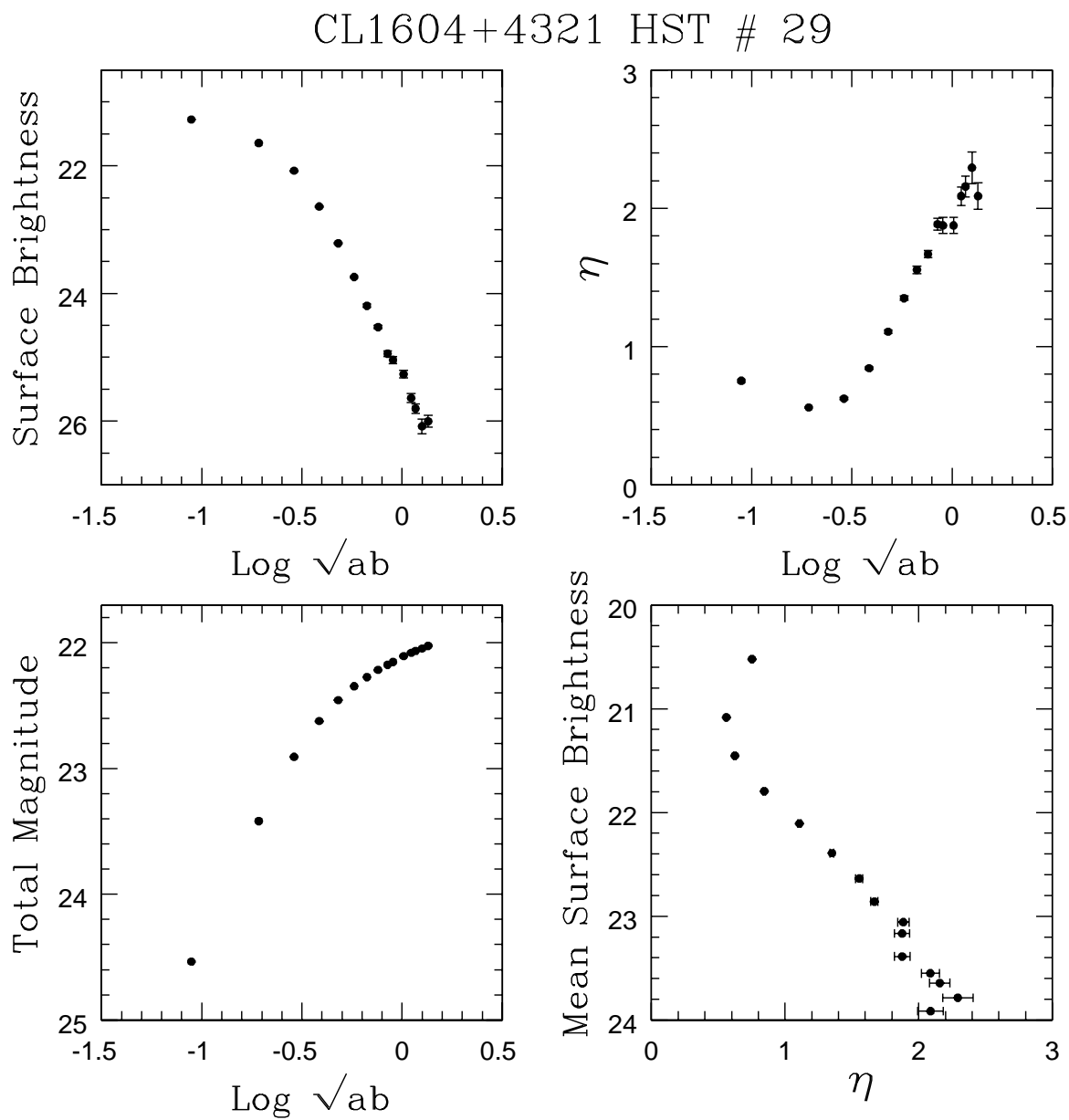}
\caption{The same as Figures 2 \& 4 but for galaxy HST \#29 in 
Cl 1604+4321. The photometry is measured in the HST $F702W$
photometric system.}
\end{figure}

\begin{figure}
\plotone{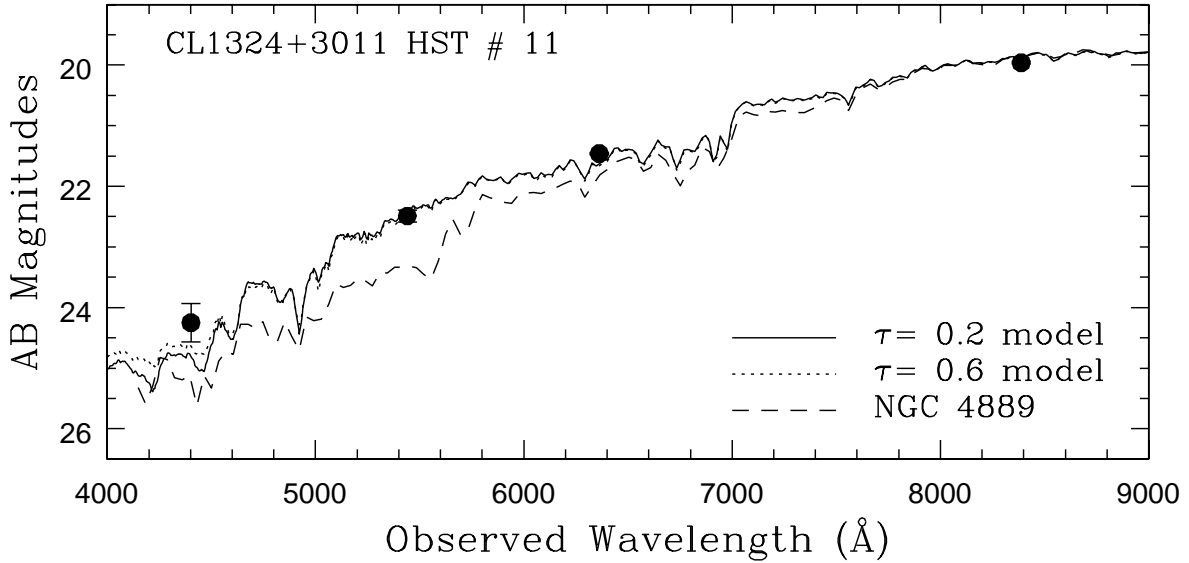}
\caption{The best-fit Bruzual \& Charlot (1993) models to the observed
broad-band ($BVRI$) AB magnitudes (shown by the closed circles) of
galaxy HST \#11 in Cl 1324+3011. The best-fit models have been
redshifted to the appropriate observed wavelength by $(1+z)$, where $z
= 0.7580$ for HST \# 11 (PLO01). The $\tau = 0.2$ Gyr model (solid
line) and the $\tau = 0.6$ Gyr model (dotted line) are shown. The
final best-fit spectrum is nearly identical in both cases (see \S
3). For comparison, we also show the spectrum for a nearby elliptical
galaxy, NGC 4889 (dashed line), redshifted to $z = 0.7580$ and
normalized to the $I$ band magnitude of HST \#11. This spectrum would
represent the no-evolution case, which is clearly inconsistent with
the observed data.}
\end{figure}

\end{document}